\documentclass[prb,nofootinbib,twocolumn,superscriptaddress]{revtex4} 
\usepackage{graphicx}% Include figure files
\usepackage{dcolumn}% Align table columns on decimal point
\usepackage{bm}% bold math 
\usepackage{threeparttable}
\usepackage{times}
\usepackage{mathptmx}
\usepackage{lscape}
\usepackage{natbib}
\usepackage{amsmath}
\usepackage{amssymb}
\usepackage{braket}
\usepackage{comment}
\usepackage{color}

\def\degree{\kern-.2em\r{}\kern-.3em}

\begin{document}

\title{ Orthogonal Decomposition of Discretization-Induced Transport-Information Cost\\ under Rank-Deficient Parametrizations
 }

\author{Koretaka Yuge}
\affiliation{
Department of Materials Science and Engineering,  Kyoto University, Sakyo, Kyoto 606-8501, Japan\\
}%

\begin{abstract}
{ When we consider discretization of continuous probability distributions, it inevitably induces irreversible geometric distortion of local measure on the discretized support. While such discretziation-induced distortion is extrinsic to information geometry (IG) alone, we recently demonstrate that the discretization cost can be naturally characterized by the standard Kullback-Leibler (KL) divergence between continuous distributions as expectation of their infinitesimal parameter variations. The framework is based on the correspondence between optimal transport (OT) and IG, primarily requring the selected parameters directly identifiable with support coordinates. The present work extends the framework to more generalized parametrization $\theta$, particularly the Jacobian between $\theta$ and support coordinates is rank-deficient, which generally results in breaking down the interpretation of the discretization-induced costs as information-geometric quantities. To address the problem, we here introduce an orthogonal decomposition of the second-moment tensor onto linear subspace for the covariance matrices generated by parameter fluctuations, based on Frobenius projection. The decomposition naturally separates the discretization cost into observable and unobservable components relative to the chosen parametrization. The present formulation provides a geometric framework for analyzing partial observability of discretization-induced transport-information costs. In particular, we show that the cross-interference cost vanishes identically when the parametrization projection commutes with the Fisher information metric or second moment structure of discretization, establishing that this term quantifies the geometric mismatch between the chosen parametrization and the intrinsic distinguishability or discretization alignment. The present framework thus clarifies the role of parametrization-dependent information loss.
}
\end{abstract}

\maketitle

\section{Introduction}
Discretization of continuous probability distribution $P$ generally induces irreversible geometric distortions. While such discretization effect is not fully characterized by information geometry (IG) alone due to support incompatibility between continuous and discrete distributions, our recent study of the so-called ``UCN'' demonstrates that the discretization cost can be naturally characterized by the standard KL divergence between continuous distributions associated with their parameter variations. The framework is based on the correpondence between optimal transport (OT) and IG in the vanishing discretization scale $d\to 0$, employing the pullback of 2-Wasserstein transport cost for the discretization onto indistinguishability on statistical manifold.\cite{amari} 

To ensure the transport-information correspondence for the discretization cost, the selected parametes $\xi$ on statistical manifold should be identifiable with support coordinates $x$, more precisely,
\begin{eqnarray}
\label{eq:xxi}
^{\forall} A\in \textrm{GL}\left(f\right), \ x\mapsto Ax \ \Rightarrow \ \xi \mapsto A\xi.
\end{eqnarray}
In such cases, we have shown that the discretization cost is explicitly given by\cite{ucn}
\begin{eqnarray}
\label{eq:cost}
\textrm{Tr}\left(M\Omega\right) = 2\mathbb{E}_{\rho\left(\delta\xi\right)} D\left( P_{\xi}:P_{\xi+\delta\xi} \right),
\end{eqnarray}
where $M$ denotes second-moment matrix of the discretization cell (as any bounded convex set), $\Omega$ the Fisher metric associated with support-coordinate variations, $P_{\xi}$ and $P_{\xi+\delta \xi}$ are nearby continuous distrbutions, and $\rho\left(\delta\xi\right)$ is the probability density for infinitesimal parameter variation $\delta\xi$. Eq.~\eqref{eq:cost} represents that the discretization cost measured by OT (l.h.s.) can be identified with twice the local indistinguishability between nearby continuous distributions as expected KL divergence under $\rho$ (r.h.s.). 

However, in general parametrizations, the mapping between variations in selected parameter $\theta$ and physical displacements on sample space is not necessarily invertible: Consequently, the associated Jacobian can become rank-deficient, implying that information about certain physical displacements cannot be represented through parameter fluctuations. In such cases, the structure for OT-based discretization cost cannot generally be reconstructed solely from the expected KL divergence as in Eq.~\eqref{eq:cost}.

The purpose of the present work is to formulate a natural decomposition of discretization-induced transport-information costs under such rank-deficient parametrizations. Rather than enforcing a complete transport-information correspondence, we decompose the physical second-moment structure into sectors observable and unobservable through the selected parametrization. 
The present framework relies on the fact that discretization is regarded primarily as an operation acting on the underlying sample space, inducing a transport distortion of the local measure. The corresponding information-geometric representation of Eq.~\eqref{eq:cost} is therefore viewed as a secondary description, obtained by projecting this geometric distortion onto the statistical manifold through the transport-information correspondence.

It should be emphasized that the present decomposition is not introduced at the level of Fisher information itself. Rather, it concerns the discretization cost established in the UCN framework, where transport-induced geometric distortion on the sample space admits an information-theoretic representation. The observable and unobservable sectors identified below therefore refer to the representability of this discretization cost under a given parametrization, rather than to the existence of the cost itself.
The geometric structure underlying the decomposition is developed in the following sections.

\section{Concept and Derivation}
\subsection{Discretization Geometry}
Let us consider a $f$-dimensional continuous support space with coordinates
\begin{eqnarray}
x \in \mathbb	{R}^{f}.
\end{eqnarray}
Then we discretize the support space with translation of any bounded convex set $\omega\subset \mathbb{R}^{f}$. 
For this discretization, we introduce a corresponding second moment matrix as
\begin{eqnarray}
M_{x} = \frac{1}{V_{\omega}} \int_{\omega} uu^{\mathrm{T}} du,
\end{eqnarray}
where the coordinate $u$ is chosen such that its mean over $\omega$ vanishes. Under this definition, the followings are satisfied:
\begin{eqnarray}
M_{x} \in \textrm{Sym}\left(f\right), \ M_{x} \succeq 0.
\end{eqnarray}
Under these preparations, the associated discretization cost at $d\to 0$ is given by $\textrm{Tr}\left(M_{x}\Omega\right)$ as seen, where $\Omega$ denotes Fisher metric with parameters identifiable with support coordinate given by Eq.~\eqref{eq:xxi}.
In the present formulation, $M_{x}$ is regarded as the fundamental geometric object associated with discretization itself.

\subsection{General Parametrization}
Now consider a general parametrization for $f$-dimensional continuous distribution $P$ of interest on statistical manifold:
\begin{eqnarray}
\theta \in \mathbb{R}^{m},
\end{eqnarray}
with local displacement map on support space of
\begin{eqnarray}
x = \zeta\left(\theta\right).
\end{eqnarray}
Then the Jacobian is given by
\begin{eqnarray}
J = \frac{\partial x}{\partial \theta} \in \mathbb{R}^{f\times m}.
\end{eqnarray}

In order to hold on the transport-information correspondence for discretization cost, Eq.~\eqref{eq:cost}, covariance matrix $\Lambda$ for $\rho\left(\delta\xi\right)$ should be proportional to the second moment structure 
\begin{eqnarray}
\Lambda \propto M_{x}.
\end{eqnarray}
Meanwhile, when we construct covariance for support space from the selected parameter covariance $\Lambda_{\theta}$, it is given by
\begin{eqnarray}
\Lambda'_{x} = J \Lambda_{\theta} J^{\textrm{T}}.
\end{eqnarray}
We then define the subspace including the obserbvable covariance as
\begin{eqnarray}
S_{J} = \left\{JYJ^{\textrm{T}} \mid Y\in \textrm{Mat}\left(m\right)  \right\}.
\end{eqnarray}
When $J$ is rank-deficient, it is clear that 
\begin{eqnarray}
S_{J} \subsetneq \textrm{Mat}\left(f\right),
\end{eqnarray}
and therefore, not all second moment matrix on support space can be represented through general parameter fluctuations. 

\subsection{Orthogonal Projection for Discretization Cost}
For parametrization with rank-deficient case, the problem comes from the fact that full information about covariance for support space cannot be constructed from selected parameter covariance, indicating that transport-information correspondence of Eq.~\eqref{eq:cost} breaks down under this parameter. Meanwhile, since the discretization is in nature  attributed to the geometric distortion on support space, the transport cost of $\textrm{Tr}\left(M_{x}\Omega\right)$ should be invariant regardless of the breaking down of the correspondence. This fact certainly indicates that the discretization cost $\textrm{Tr}\left(M_{x}\Omega\right)$ should be decomposed into observable and unobservable counterpart for the selected parameter fluctuations. 

Here we employ the orthogonal projection of the covariance $\Lambda_{x}$ in terms observability in $\theta$ space under appropriate inner product, namely,  
\begin{eqnarray}
\label{eq:dc}
\Lambda_{x} = \Lambda_{\parallel} + \Lambda_{\perp} + \Lambda_{\textrm{c}},
\end{eqnarray} 
where the first term of the r.h.s. corresponds to the projected covariance (i.e., observable part), the second term the orthogonal conterpart (unobservable part), and the third is the cross term reflecting their interference. 
Eq.~\eqref{eq:dc} requires the definition of inner product $H$ for
\begin{eqnarray}
^{\forall}X\in\textrm{Mat}\left(m\right), \ \Braket{\Lambda_{x}-\Lambda_{\parallel},JXJ^{\textrm{T}}}_{H} =0.
\end{eqnarray}
We here adopt Frobenius inner product,\cite{FB} yielding to
\begin{eqnarray}
\Braket{\Lambda_{x}-\Lambda_{\parallel},JXJ^{\textrm{T}}}_{F} &=& \textrm{Tr}\left[\left(\Lambda_{x} - \Lambda_{\parallel}\right) JXJ^{\textrm{T}}\right]\nonumber \\
&=& \textrm{Tr}\left[J^{\textrm{T}}\left(\Lambda_{x} - \Lambda_{\parallel}\right)J X\right]= 0.
\end{eqnarray}
Since $X$ can be chosen arbitrarily, the relation
\begin{eqnarray}
^{\forall}X \in \textrm{Mat}\left(m\right), \quad \textrm{Tr}\left(AX\right)=0
\end{eqnarray}
implies
\begin{eqnarray}
A=0.
\end{eqnarray}
Therefore, we obtain
\begin{eqnarray}
J^{\textrm{T}} \left(\Lambda_{x} - \Lambda_{\parallel}\right)J =0.
\end{eqnarray}
Therefore, the Frobenius inner product provides appropriate condition of
\begin{eqnarray}
\label{eq:cond}
J^{\textrm{T}}\Lambda_{x} J = J^{\textrm{T}} \Lambda_{\parallel} J.
\end{eqnarray}
Eq.~\eqref{eq:cond} corresponds to address the observability from $\theta$ space, because Eq.~\eqref{eq:cond} leads to 
\begin{eqnarray}
^{\forall}\delta\theta\in\mathbb{R}^{m}, \left(J\delta\theta\right)^{\textrm{T}}\Lambda_{x} J \delta\theta = \left(J\delta\theta\right)^{\textrm{T}}\Lambda_{\parallel} J \delta\theta,
\end{eqnarray}
which means that $\Lambda_{x}$ and $\Lambda_{\parallel}$ are indistinguishable in terms of their quadratic form, w.r.t. the $\theta$-variation induced restricted physical displacement, $\delta x' = J \delta\theta$.

Under this inner product, we first introduce the operator $P_{J}$  as
\begin{eqnarray}
P_{J} = J\left(J^{\textrm{T}}J\right)^{+}J^{\textrm{T}},
\end{eqnarray}
where superscript $^{+}$ denotes Moore-Penrose inverse.\cite{mp} 
Then we can construct the orthogonal projection map
\begin{eqnarray}
\Phi: \ \textrm{Mat}\left(f\right) \mapsto  S_{J}
\end{eqnarray}
by
\begin{eqnarray}
\Phi\left(\Lambda\right) = P_{J}\Lambda P_{J}.
\end{eqnarray}
Indeed, since $P^{2}_{J}=P_{J}$ and $P_{J}^{\textrm{T}}=P_{J}$, the map $\Phi=P_{J}\Lambda P_{J}$ gives the Frobenius-orthogonal projection onto $S_{J}$.
With these preparations, we can straightforwardly obtain the observable component of the covariance information as
\begin{eqnarray}
\Lambda_{\parallel} = P_{J} \Lambda_{x} P_{J},
\end{eqnarray}
while the unobservable component is defined as 
\begin{eqnarray}
\Lambda_{\perp} = \left(I - P_{J}\right) \Lambda_{x} \left(I-P_{J}\right),
\end{eqnarray}
and the residual is the inteference cross-term of 
\begin{eqnarray}
\Lambda_{\textrm{c}} = P_J \Lambda_{x} \left(I-P_J\right) + \left(I-P_J\right) \Lambda_{x} P_J.
\end{eqnarray}
These certainly satisfy the orthogonality:
\begin{eqnarray}
\Braket{\Lambda_{\parallel},\Lambda_{\perp}}_{F} &=& 0 \nonumber \\
\Braket{\Lambda_{\parallel},\Lambda_{\textrm{c}}}_{F} &=& 0 \nonumber \\
\Braket{\Lambda_{\perp},\Lambda_{\textrm{c}}}_{F} &=& 0,
\end{eqnarray}
and the positive semidefiniteness (PSD) condition for observable and unobservable part:
\begin{eqnarray}
\Lambda_{\parallel} \succeq 0, \ \Lambda_{\perp} \succeq 0,
\end{eqnarray}
while the inteference part $\Lambda_{\textrm{c}}$ is indefinite. 
We can briefly confirm the PSD condition of the present projection, to retain the covarince structure. For instance, since $\Lambda_{x}\succeq 0$, the projected matrix also satisfies 
\begin{eqnarray}
\Lambda_{\parallel} = P_{J} \Lambda_{x} P_{J} \succeq 0.
\end{eqnarray}
Indeed, 
\begin{eqnarray}
^{\forall}v \in \mathbb{R}^{f}, \ v^{\textrm{T}} \Lambda_{\parallel} v = \left(P_{J}v\right)^{\textrm{T}} \Lambda_{x} \left(P_{J }v\right) \ge 0.
\end{eqnarray}

From these discussions, therefore, the discretization cost admits the following decomposition:
\begin{eqnarray}
\label{eq:dcp}
\textrm{Tr}\left(M_{x}\Omega\right) = \textrm{Tr}\left(M_{\parallel}\Omega\right) + \textrm{Tr}\left(M_{\perp}\Omega\right) + \textrm{Tr}\left(M_{\textrm{c}}\Omega\right),
\end{eqnarray}
where $M_{\parallel}\propto \Lambda_{\parallel}$, $M_{\perp}\propto \Lambda_{\perp}$ and $M_{\textrm{c}}\propto \Lambda_{\textrm{c}}$with setting their commom proportional constant.
The first term in r.h.s. represents the cost reconstructable from the selected parameter fluctuations, the second term corresponds to physically existing contributions to the cost invisible to such parametrization, and the third term the nontrivial interference contributions between observable and unobservable sector.  Thus, under rank-deficient parametrizations, the failure of transport-information correspondence is naturally interpreted as a geometric loss of observable transport directions.

Since $\Omega\succeq 0$, we can confirm the nonnegative character of the observable and unobservable cost: 
\begin{eqnarray}
\textrm{Tr}\left(M_{\parallel}\Omega\right) \ge 0, \ \textrm{Tr}\left(M_{\perp}\Omega\right) \ge 0,
\end{eqnarray}
while the interference cost is indefinite. 
However, we can easily show that when 
\begin{eqnarray}
\left[P_{J},\Omega\right] = 0
\end{eqnarray}
or
\begin{eqnarray}
\left[ P_{J}, M \right] = 0
\end{eqnarray}
is satisfied, 
\begin{eqnarray}
\textrm{Tr}\left(M_{\textrm{c}}\Omega\right) = 0
\end{eqnarray}
should always hold on. 
This mathematical structure implies that when $P_{J}$ and $\Omega$ and/or $P_{J}$ and $M$ is commutative, the interference contribution vanishes identically. This geometric feature ensures that even when an observable direction is transformed or evolved by the distinguishability or discretization structure, the resulting information variation remains entirely confined within the observable sector.

\subsection{Cost Bounds with the Geometric Interpretation} 
\begin{figure}[h]
\begin{center}
\includegraphics[width=0.77\linewidth]{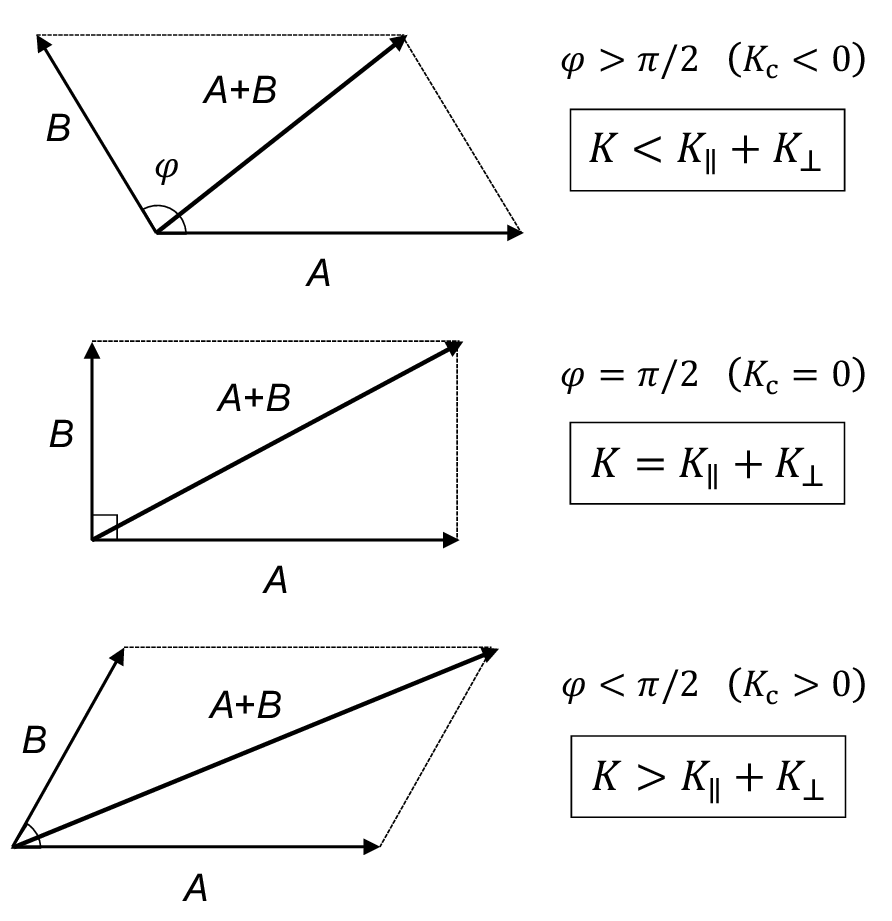}
\caption{ Geometric interpretation of total, projected, orthogonal, and interference costs in the Frobenius inner-product space. The sign of the interference cost $K_{\textrm{c}}$ is determined by the angle $\phi$ between matrices $A$ and $B$.}
\label{fig:gm}
\end{center}
\end{figure}

Based on the above discussion, we here address how the total discretization cost $K=\textrm{Tr}\left( M_{x}\Omega \right)$ and the decomposed costs of $K_{\parallel}=\textrm{Tr}\left( M_{\parallel}\Omega \right)$, $K_{\perp}=\textrm{Tr}\left( M_{\perp}\Omega \right)$ and $K_{\textrm{c}}=\textrm{Tr}\left( M_{\textrm{c}}\Omega \right)$ are mutually bounded.  
For this purpose, we first introduce the following matrix products:
\begin{eqnarray}
A &=& M_{x}^{1/2}P_{J}\Omega^{1/2} \nonumber \\
B &=& M_{x} ^{1/2}\left( I - P_{J} \right)\Omega^{1/2}.
\end{eqnarray}
With these definitions, we can find that the decomposed costs admits the following expressions:
\begin{eqnarray}
K_{\parallel} &=& \textrm{Tr}\left( P_{J}M_{x}P_{J}\Omega \right) = \textrm{Tr}\left( M_{x}^{1/2}P_{J}\Omega P_{J}M_{x}^{1/2} \right) = \left\| A\right\|^{2}_{F} \nonumber\\
K_{\perp} &=& \textrm{Tr}\left( M_{x}^{1/2}\left(I- P_{J} \right)\Omega \left(I- P_{J} \right)M_{x}^{1/2} \right) = \left\|B\right\|^{2}_{F} \nonumber \\
K_{\textrm{c}} &=& 2\Braket{A, B}_{F}.
\end{eqnarray}
From the above equations, the Cauchy-Schwarz inequality
\begin{eqnarray}
\left| \Braket{A,B}_{F} \right| \le \left\|A\right\|_{F} \left\|B\right\|_{F},
\end{eqnarray}
and Eq.~\eqref{eq:dcp}, we obtain the  bounds for interference cost:
\begin{eqnarray}
-2\sqrt{K_{\parallel}K_{\perp}} \le K_{\textrm{c}} \le 2\sqrt{K_{\parallel}K_{\perp}}
\end{eqnarray}
and the bounds for total cost:
\begin{eqnarray}
\left( \sqrt{K_{\parallel}} - \sqrt{K_{\perp}} \right)^{2} \le K \le \left( \sqrt{K_{\parallel}} + \sqrt{K_{\perp}} \right)^{2}.
\end{eqnarray}
Since the total cost can be rewritten as $K=\left\|A+B\right\|^{2}_{F}$, the derived bounds among total and the decomposed cost can be geometrically interpreted based on the parallelogram relationship shown in Fig.~\ref{fig:gm}, where we define the angle $\phi$ as 
\begin{eqnarray}
\cos \phi = \frac{\Braket{A,B}_{F}}{ \left\|A\right\|_{F}\left\|B\right\|_{F}}.
\end{eqnarray}
Under this definition, the interference cost can be rewritten as 
\begin{eqnarray}
K_{\textrm{c}} = 2\sqrt{K_{\parallel}K_{\perp}}\cos \phi. 
\end{eqnarray}
Therefore, (i) the sign of the interference cost $K_{\textrm{c}}$ is determined by the angle $\phi$ between matrices $A$ and $B$, and (ii) the amplitude of the interference is characterized by the difference in squared diagonal lengths of the parallelogram.  
From these geometric depictions, we can clearly see that acute ($\phi < \pi/2$), orthogonal ($\phi=\pi/2$), and obtuse ($\phi>\pi/2$) relations between $A$ and $B$ respectively corresponds to constructive, vanishing, and destructive interference of discretization costs.

\subsection{Geometry of Cost Observability in Rank-Deficient Linear Maps }
To concretely see the geometric insights for the proposed costs, we here present a minimal yet representative, illustrative case: Discretizaion of a two-dimensional continuous distribution ($f=2$) observed from a single parametrization $\eta$ ($m=1$) through linear transform of the support coordinates. This simple linear reduction reasonably captures how the full transport geometry is partitioned into observable, unobservable and interference  sectors. 

Consider a two-dimensional continuous distribution of support coordinates $x=\left( x_{1}, x_{2} \right)$, $P\left( x \right)$, and its Fisher metric under appropriate parameters identifiable with $x$,
\begin{eqnarray}
\Omega =
\begin{pmatrix}
\omega_{1} & W \\
W & \omega_{2}
\end{pmatrix}.
\end{eqnarray}
$P$ is then discretized by translation of any convex bounded set, whose second moment matrix is given by
\begin{eqnarray}
M=
\begin{pmatrix}
m_{1} & L\\
L & m_{2}
\end{pmatrix}.
\end{eqnarray}
When we consider a rank-deficient parametrization $\eta$ from $f=2$ to $m=1$ through linear map, $\eta$ can always be expressed as the following rotational form up to overall constant factor:
\begin{eqnarray}
\label{eq:eta}
\eta = x_{1} \cos\alpha + x_{2} \sin\alpha.
\end{eqnarray}
The corresponding Jacobian and its orthogonal complement can be respectively given by
\begin{eqnarray}
J = \dfrac{\partial x}{\partial \eta } = 
\begin{pmatrix}
\cos \alpha \\
\sin  \alpha
\end{pmatrix}, \
J_{\perp} = 
\begin{pmatrix}
-\sin \alpha \\
\cos\alpha
\end{pmatrix}.
\end{eqnarray}
For the present setup of $\eta$ in Eq.~\eqref{eq:eta}, 
\begin{eqnarray}
\left( J^{\textrm{T}}J \right)^{+} = 1
\end{eqnarray}
holds on.
Therefore, by using the relationships
\begin{eqnarray}
P_{J} = JJ^{\textrm{T}},\ \left( I-P_{J} \right)=J_{\perp}J_{\perp}^{\textrm{T}},
\end{eqnarray}
we can rewrite the interference cost as
\begin{eqnarray}
K_{\textrm{c}} &=& \textrm{Tr}\left[ \left( J^{\textrm{T}}MJ_{\perp} \right)\left( J_{\perp}^{\textrm{T}}\Omega J \right) + \left( J_{\perp}^{\textrm{T}}MJ \right) \left( J^{\textrm{T}}\Omega J_{\perp} \right) \right] \nonumber \\
&=& 2\left( J^{\textrm{T}}MJ_{\perp} \right)\left( J^{\textrm{T}}\Omega J_{\perp} \right).
\end{eqnarray}
Now consider an eigendecomposition of $M$ in the following form:
\begin{eqnarray}
M &=& Q M_{\textrm{D}} Q^{\textrm{T}}
\end{eqnarray}
with
\begin{eqnarray}
M_{\textrm{D}} =
\begin{pmatrix}
\lambda_{+} & 0\\
0 &\lambda_{-}
\end{pmatrix} ,\quad
Q =
\begin{pmatrix}
\cos\alpha_{M} & -\sin\alpha_{M} \\
\sin\alpha_{M} & \cos\alpha_{M}
\end{pmatrix},
\end{eqnarray}
where $\lambda_{+}\ge \lambda_{-}$, and 
\begin{eqnarray}
\tan\left( 2\alpha_{M} \right) = \dfrac{2L}{m_{1}-m_{2} }.
\end{eqnarray}
In a similar fashion, we perform an eigendecomposition of the Fisher metric as $\Omega=Q'\Omega_{\textrm{D}}Q'^{\textrm{T}}$, with $\textrm{diag}\left( \Omega_{\textrm{D}} \right)=\left\{ \mu_{+},\mu_{-} \right\}$, and the rotaional matrix $Q'$ characterized by the following angle $\alpha_{\Omega}$:
\begin{eqnarray}
\tan\left( 2\alpha_{\Omega} \right) = \dfrac{2W}{\omega_{1}-\omega_{2} }.
\end{eqnarray}
The angles $\alpha_{M}$ and $\alpha_{\Omega}$ denote the directions of the eigenvectors associated with the largest eigenvalues $\lambda_{+}$ and $\mu_{+}$, respectively. We also introduce the followings:
\begin{eqnarray}
&&\Delta\lambda = \dfrac{\lambda_{+}-\lambda_{-}}{2}, \ \Delta\mu = \dfrac{\mu_{+}-\mu_{-}}{2} \nonumber \\
&& \lambda_{0} = \dfrac{\lambda_{+}+\lambda_{-}}{2}, \ \mu_{0} = \dfrac{\mu_{+}+\mu_{-}}{2}.
\end{eqnarray}

Under these preparations, we can straightforwardly express the interference cost as
\begin{eqnarray}
\label{eq:kc}
K_{\textrm{c}} = 2\Delta\lambda\Delta\mu \sin\left( 2\alpha_{M} - 2\alpha \right) \sin\left( 2\alpha_{\Omega} - 2\alpha \right).
\end{eqnarray}
Eq.~\eqref{eq:kc} certainly indicates that amplitude of the interference cost is dominated by anisotropy of the discretization and distinguishable geometry of $\Delta\lambda$ and $\Delta\mu$, while its sign is determined from the geometric alignment of the observing parameter $\eta$ w.r.t. the principal direction of the discretization and distinguishable geometry: When observing direction lies between the two principal directions, the interference provides destructive contribution $K_{\textrm{c}}<0$, and vice versa. The commutative condition $\left[ P_{J},\Omega \right]=0$ corresponds to $\alpha_{\Omega}-\alpha=0,\ \pi/2$, which is a sufficient condition of $K_{\textrm{c}}=0$.

In a similar fashion to the interference cost, we can rewrite the observable and unobservable costs as
\begin{eqnarray}
K_{\parallel} = \dfrac{1}{2}\left( K-K_{\textrm{c}} \right) + H \nonumber \\
K_{\perp} = \dfrac{1}{2}\left( K-K_{\textrm{c}} \right) - H, 
\end{eqnarray}
where
\begin{eqnarray}
H =  \mu_{0}\Delta\lambda\cos\left( 2\alpha_{M}-2\alpha \right) + \lambda_{0}\Delta\mu\cos\left( 2\alpha_{\Omega}-2\alpha \right). 
\end{eqnarray}
The imbalance between observable and unobservable costs is governed by two additive contributions:
One arising from the anisotropy of the discretization geometry weighted by the average distinguishability $\mu_{0}$,
and the other arising from the anisotropy of the distinguishability geometry weighted by the average discretization scale $\lambda_{0}$.  Their signs are determined by the $\pi/2$-rotaional alignment of the observing parameter w.r.t. their principal directions. 

\subsection*{Summary}
The present decomposition clarifies that rank-deficient parametrizations do not necessarily invalidate transport geometry itself. Rather, they restrict the subset of transport-induced fluctuations observable through parameter variations.
This viewpoint suggests that information geometry under general parametrizations should not always be expected to reproduce the full physical discretization geometry. Instead, the parametrization defines an observable transport sector embedded within the full covariance geometry of the support space.
The present formulation may provide a useful framework for analyzing partial observability, coarse-graining, and hidden transport directions in statistical manifolds.

\section{Conclusions}
We introduced an orthogonal decomposition framework for discretization-induced transport-information costs under rank-deficient parametrizations. By projecting covariance structures on support space onto those generated by parameter fluctuations, the cost naturally separates into observable, unobservable, and cross-interference sectors. 
Crucially, we have demonstrated that under the commutativity between the projection operator and the Fisher metric or second moment structure of discretization cell, the interference cost vanishes identically because the information variation remains entirely confined within the observable sector. 

This formulation provides a geometric interpretation of the breakdown of transport-information correspondence and establishes a mathematically consistent framework for analyzing parametrization-dependent observability of transport-induced discretization costs.

\section{Acknowledgement}
This work was supported by JSPS KAKENHI Grant Number 23K04359 and Research Grant from Hitachi Metals$\cdot$Materials Science Foundation.

\end{document}